\documentclass[a4paper,fleqn,usenatbib]{mnras}

\usepackage{mathptmx}

\usepackage[T1]{fontenc}
\usepackage{ae,aecompl}

\include{definitions}

\defcitealias{riess_2.4_2016}{R16}
\newcommand{\Riess}{\citetalias[][]{riess_2.4_2016}}

\usepackage{graphicx}	
\usepackage{amsmath}	
\usepackage{amssymb}

\newcommand{\lcdm}{$\Lambda$CDM}

\mathchardef\mathhyphen="2D
\title[Hubble Constant and Cepheid Calibration Modeling Choices]
{Insensitivity of The Distance Ladder Hubble Constant Determination to Cepheid Calibration Modeling Choices}
\author[Follin et. al.]
  {B.~Follin,$^1$\thanks{aarrasmith@ucdavis.edu}
  L.~Knox$^1$\\
  $^1$Department of Physics, University of California, Davis\\
  One Shields Ave, Davis, CA 95616\\}
\date{July 2017}

\begin{document}
\maketitle
\begin{abstract}
Recent determination of the Hubble constant via Cepheid-calibrated supernovae by \citet{riess_2.4_2016} (R16) find $\sim 3\sigma$ tension with inferences based on cosmic microwave background temperature and polarization measurements from $Planck$. 
This tension could be an indication of inadequacies in the concordance $\Lambda$CDM model. 
Here we investigate the possibility that the discrepancy could instead be due to systematic bias or uncertainty in the Cepheid calibration step of the distance ladder measurement by R16. We consider variations in total-to-selective extinction of Cepheid flux as a function of line-of-sight, hidden structure in the period-luminosity relationship, and potentially different intrinsic color distributions of Cepheids as a function of host galaxy. Considering all potential sources of error, our final determination of $H_0 = 73.3 \pm 1.7~{\rm km/s/Mpc}$ (not including systematic errors from the treatment of geometric distances or Type Ia Supernovae) shows remarkable robustness and agreement with R16. We conclude systematics from the modeling of Cepheid photometry, including Cepheid selection criteria, cannot explain the observed tension between Cepheid-variable and CMB-based inferences of the Hubble constant. Considering a `model-independent' approach to relating Cepheids in galaxies with known distances to Cepheids in galaxies hosting a Type Ia supernova and finding agreement with the R16 result, we conclude no generalization of the model relating anchor and host Cepheid magnitude measurements can introduce significant bias in the $H_0$ inference.
\end{abstract}

\section{Introduction}
The standard cosmological paradigm of a universe dominated by standard model particles, cold dark matter, and a cosmological constant ($\Lambda$CDM), with adiabatic, nearly scale-invariant initial density perturbations, has remarkable and continued success in modeling a host of cosmological observations, including the Cosmic Microwave Background (CMB) temperature and polarization anisotropies \citep{planck_collaboration_planck_2016-1}, the Baryon-Acoustic Oscillation (BAO) feature in the galaxy number power spectrum \citep{ross_clustering_2016}, lensing of CMB photons from late-time matter inhomogeneities \citep{planck_collaboration_planck_2016-1}, and the slope of the apparent-magnitude to redshift relation of type Ia supernovae (SNIa) \citep{rest_cosmological_2014}. With this success, the degrees of freedom of the \lcdm{} model, often parameterized as the densities of matter ($\omega_{\rm m}$) and cold dark matter ($\omega_{\rm c}$), the angular size of the acoustic horizon at recombination $\theta_s$, the optical depth to recombination $\tau$, the amplitude ($A_s$) and tilt ($n_s$) of the primordial fluctuation power spectrum, and the value of the cosmological constant $\Lambda$, have been determined to $\sim 1\%$ scale accuracy. This exquisite precision, mostly driven by the inceasingly precise measurements of the CMB from $Planck$ \citep{planck_collaboration_planck_2016-1}, allows for the tight prediction (within the model) of quantities important to the cosmic evolution at any epoch. Consequently, though the above probes are not directly sensitive to the current Hubble rate, $H_0$, under the \lcdm{} model CMB measurements place a precise constraint of $H_0 = 66.93 \pm 0.62 {\rm \, km/s/Mpc}$ \citep{planck_collaboration_planck_2016-1}, with this value completely consistent with constraints imposed by the other cosmological probes above \citep{planck_collaboration_planck_2016}, \citep{aubourg_cosmological_2015}.

Complementing these cosmological probes mentioned above, considerable effort has been made on more direct astrophysical measurements of the current Hubble rate. Objects obtain a redshift due to cosmic expansion given approximately by 
\begin{equation}
\label{eq: hubble rate}
z_c = H_0 d/c.
\end{equation}
 A simultaneous measurement of $z_c$ and the proper distance $d$ for an object would in principle allow for a direct measurement of $H_0$. However, for $z_c$ to be a considerable contribution to the total measured redshift $z$, the object must be sufficiently far away that direct determinations of distance (through, e.g. parallax) has hitherto been impossible. In place of a direct measurement, a laddering approach has been developed that calibrates the absolute magnitude of type IA supernovae (SNIa) in resolvable galaxies with Cepheid variable stars, whose own absolute magnitude is obtained either through comparison to direct parallax of Cepheids in the Milky Way, or a host of astronomically-estimated distance measures to nearby galaxies. A recent analysis by \citet{riess_2.4_2016} (hereafter R16) uses a combination of Cepheids in 22 nearby galaxies and the Milky Way, 19 of which host type 1A supernovae, to obtain $H_0 = 73.24 \pm 1.74 {\rm \, km/s/Mpc}$. This result is formally in $ 3.4 \sigma$ tension with the \lcdm{} prediction conditioned on the CMB temperature and polarization spectra.

This discrepancy has several potential causes: an unlikely statistical fluke, a systematic bias in the local astronomical measurement of \Riess{}{} or in the CMB measurements used to condition \lcdm{}, or a breakdown of the \lcdm{} model. The last of these options has been considered elsewhere in the literature, in e.g. \citet{chacko_partially_2016}, \citet{karwal_early_2016}, \citet{di_valentino_reconciling_2016}, and \citet{bernal_trouble_2016}. The simplest physically motivated extensions to \lcdm, however, fail to completely relieve the tension \citep{hou_constraints_2014,planck_collaboration_planck_2016}. Perhaps most effective extension is a variable effective number of neutrino species, $N_{\rm eff}$, though varying $N_{\rm eff}$ only relaxes the tension to $2.2\sigma$. Varying the equation-of-state parameter $w$, another common phenomenological change that brings the CMB inference of $H_0$ lower, requires $w < -1$. Such a value is difficult to understand theoretically \citep{carroll_can_2003}. If the tension is to be completely explained through new physics it will require a significant departure from \lcdm{} in an unexpected manner. 

Such a claim will require extraordinary evidence, so there has been recent interest in revisiting the \Riess{} analysis and exploring the impact of dropping various assumptions made therein. In lieu of fixed error bars, \citet{cardona_determining_2017} elevate uncertainties in Cepheid photometry to parameters to be jointly estimated together with $H_0$. They find the tension slightly reduced to $3.1\sigma$. \citet{feeney_clarifying_2017} generalize the assumed Gaussian distributions of uncertainty to $t$ distributions and find odds against \lcdm{} (under the assumption of no systematics) at 65:1.

In this paper we investigate the possibility of an underestimated systematic in the Cepheid calibration central to the local $H_0$ determination. While acknowledging the possibilities of errors elsewhere, for the remainder of this paper we will assume the validity of \lcdm, and that the $Planck$ determination of the \lcdm{} free parameters is generally accurate. 
Some justification for this latter conjecture comes from the multiplicity of independent measurements at high redshifts, including $Planck$ \citep{planck_collaboration_planck_2016-1}, South Pole Telescope \citep{hou_comparison_2017,kevin}, and Atacama Cosmology Telescope \citep{louis_atacama_2014} temperature and polarization anisotropy measurements and CMB lensing estimates from the CMB 4-point function as measured by $Planck$. These measurements have shown remarkable consistency under the \lcdm{} hypothesis \citep{marius,kevin}, and a systematic bias common to all seems a remote possibility, although see \citet{addison_quantifying_2016} for a dissenting viewpoint.

Alternative local measurements likewise agree with the \citet{riess_2.4_2016} measurement; a time-delay measurement by the H$0$LiCOW collaboration \citep{bonvin_h0licow_2017} find $72.8 \pm 2.4$ km/s/Mpc through analysis on three multiply imaged Quasar systems and with uniform prior on $H_0$ and fixed value of $\Omega_{m} = 0.32$, in agreement with \Riess{} and in mild ($2.5\sigma$) tension with the $Planck$ value. 
Regardless, as the tension between local and CMB-inferred measurements of the Hubble rate is most notable when comparing to the Cepheid-derived constraints of \Riess{}, we focus on examining potential systematics in this measurement. In section \ref{sec: R16 rehash}, we provide an overview of the \Riess{} measurement, and argue that a natural place to look for systematics is in the modeling of Cepheid magnitudes in the period-metallicity-color space. In section \ref{sec: new analysis} we present a generalized framework for checking the dependence of $H_0$ on potential systematics in any of the above Cepheid features, while in section \ref{sec: results} we give updated constraints on $H_0$ after relaxing the treatment of Cepheid color, Cepheid extinction along the line of sight, and the Cepheid period-magnitude relationship. We explore the remarkable consistency with the baseline treatment in \Riess{}, and construct a `model free' approach that highlights the general insensitivity of the $H_0$ inference to choices made in modeling Cepheid magnitudes. In section \ref{sec: conclusion} we discuss the state of the tension and future implications. 

\section{Determining $H_0$ from a Distance Ladder}
\label{sec: R16 rehash}
In this section we present an overview of the \Riess{} analysis, to frame our discussion of potential systematic effects. In a nutshell, the \Riess{} analysis starts with a calibration of the Cepheid period-magntitude relationship, which then enables determination of the absolute magnitude of a (standardized) SNIa. 
Given the absolute magnitude of a supernova or other cosmological object and its apparent magnitude, or equivalently its luminosity distance, one can then directly determine the local Hubble rate. In a flat universe, the luminosity distance is given by 
\begin{equation}
\label{eq: luminosity distance}
D_{\rm lum} = \left[\frac{c}{H_0} \int_0^z \frac{dz'}{h(z')}\right](1+z)
\end{equation}
with $h(z) = H(z)/H_0$ determined through the evolution of energy density predicted by \lcdm{} and constrained by cosmological probes. At sufficiently low redshifts (though sufficently far away so as to be in the Hubble flow), $h(z') \simeq 1$ and the cosmological dependence of equation \ref{eq: luminosity distance} disappears so $H_0 D_{\rm lum} = c z$. 

Unfortunately no object exists in the Hubble flow where a direct, accurate distance determination is possible; instead, standardizeable SNIa in the Hubble flow are compared to the population of SNIa in nearby host galaxies. As yet, no SNIa has been detected in a galaxy where a direct geometric distance measure has been made, so these distances must be inferred in turn through comparison to standardizeable companions in the host Galaxies whose population is also sampled in galaxies with known geometric distances. This comparison is made through Cepheid variable stars, which leads to a three-rung `laddering' of observables: a measurement of geometric distances to calibrate Cepheids, which in turn are used to calibrate SNIa in the Hubble flow. The distances to these SNIa can then be related to cosmology through equation \ref{eq: luminosity distance}.

\subsection{Geometric Measurements to Cepheid Variable Stars}
The first rung in the ladder is geometric measurements to a population of Cepheid variable stars. The most direct of these is parallax measurements of the stars themselves. However, this method is limited in scope to Cepheids inside the Milky Way. Via use of the Hubble Telescope Fine Guidance Sensor, Wide Field Camera (WFC3), and \textit{Hipparcos}, \citet{van_leeuwen_cepheid_2007} find 19 Cepheid parallaxes which give a Cepheid magnitude calibration estimate with an uncertainty of $\sim 2.5\%$. 
In addition, eight systems of detached eclipsing binary stars (DEBs) in the LMC, whose orbital dynamics allow for independent measure of both radial velocity and orbital phase, were analyzed by \citet{pietrzynski_eclipsing_2013}, providing a $2\%$ estimate of the luminosity distance to the LMC, which hosts $\sim 800$ observed Cepheid candidates. 
Finally, line-of-sight and velocity measurements \citep{humphreys_toward_2013} of a large megamaser system in NGC$4258$ (hosting $\sim 150$ Cepheid candidates) allow for a $2.6\%$ determination of the distance to the host, after accounting for system inclination and other nuisance effects.
\subsection{Cepheid Variable Star Modeling}
If a type Ia supernova existed in one or more of these galaxies, the geometric measurements above would be enough. To bridge the gap between the above anchor galaxies with known luminosity distance and SNIa host galaxies, \Riess{} catalogue around $2000$ Cepheid variable stars in the V, I, and H bands of the Hubble WFC3. 

Cepheids have an empirically-determined period-magntiude relationship with an empirically-determined width of $\sigma \simeq 0.06 \mathhyphen 0.09$ in the near-infrared. The relationship can be expressed as
\begin{align}
\label{eq: ceph period magnitude}
m_H & - R_H E(V-I)  \\
&=  \mu + M_{\rm Ceph} + b \left(\log(P) - 1 \right) + \gamma \left(\delta \log [O/H]\right), \nonumber
\end{align}
where $m_H$ is the observed apparent magnitude in the WFC$3$ H band (corrected for crowding effects), $R_H E(V-I)$ is the color correction due to extinction $A(H)$ due to dust along the line of sight, $\mu = 5 \log \left(D_{\rm lum}/{\rm Mpc}\right) + 25$ is the inferred distance modulus to the Cepheid, $M_{\rm Ceph}$ is the absolute magnitude of a period $P = 10$ day Cepheid with Milky-Way metallicity $\log [0/H]$, and $b$ and $\gamma$ are slope parameters taking into account dependence of magnitude on both metallicity and period. 

Since extinction along the line of sight is unobservable, \Riess{} make the replacement $E(V-I) \rightarrow V-I = E(V-I) + (V-I)^0$ in equation \ref{eq: ceph period magnitude}. This replacement has the potential to lead to significant biases in $H_0$ determination as the intrinsic colors are of order unity and the difference between the $Planck$ and \Riess{} values corresponds to a magnitude change of only 0.5. In the \Riess{} analysis, the replacement has no such impact as $R_H$ is fixed and the host and anchor galaxies have similar distributions of intrinsic Cepheid color. Under these conditions, the ``error'' $R_H (V-I)^0$ is simply absorbed by the nuisance term $M_{\rm Ceph}$; i.e., the same error in absolute magnitude inference is made for host as for anchor galaxies so the errors cancel out for purposes of SNIa calibration\footnote{The variation about the mean $R_H (V-I)^0$leads to increased scatter that is subdominant to photometric errors and the intrinsic width of the Cepheid instability strip.}.

 Finally, \Riess{} introduce additional freedom in equation \ref{eq: ceph period magnitude} by allowing a break in the period-magnitude relationship at $P = 10$ days through splitting the inference of the nuisance parameter $b$ into high and low period slopes $b_h$ and $b_l$, respectively. Such freedom is supported experimentally by e.g. \citet{sandage_new_2009}.

\subsection{Supernova Magnitude Determination}
The population of type Ia supernovae, whose underlying physics is expected to be consistent across samples, have lightcurves that follow roughly similar evolution over time. As such, they are standardizeable candles in the sense that the differences in these lightcurves can be parameterized by a few nuisance parameters that adjust the inferred `standardized' magnitude of a particular SNIa. This is normally done through a principal component analysis (PCA) trained on a set of known (labeled) SNIa lightcurves, who use input spectral information as a function of time to predict a standard magnitude $m_B^{0}$, meant to represent a (scaled) fiducial $B$ band magnitude at peak luminosity \citep{betoule_improved_2014}.  \Riess{} consider a set of $22$ SNIa whose host galaxy contain Cepheids, Cepheids whose distance modulii are determined by equation \ref{eq: ceph period magnitude} above.

To standardize the SNIa observed magnitudes \Riess{} use the SALT2 filter \citep{betoule_improved_2014}, which fits two empirically-determined directions of shape and color variation and a fiducial color law in order to determine a flux model $S_{\rm SN}(p, \lambda)$, where the phase $p$ parameterizes the location on the lightcurve and the wavelength $\lambda$ is the wavelength of observation. Like any machine learning predictor, an SED fitter like SALT2 may introduce bias particular to the algorithmic complexity and fiducial choices made, as variations outside of the principle directions and fiducial color law (from, say variations in dust-driven extinction over the lines of sight) may not be captured and corrected for. To check for those biases directly attributable to training strategy, \citet{mosher_cosmological_2014} simulate an array of SNIa samples, and find negligible bias on inferred cosmological parameters.

\subsection{Combining these Measurements for a Determination of $H_0$}
To combine the measurements of SNIa and Cepheids above, the laddering approach makes use of equation \ref{eq: ceph period magnitude}, along with the similar relationship for SNIa,
\begin{equation}
\label{eq: supernovae magnitude}
m_B^0 = \mu + M_0^{\rm SN}.
\end{equation}
Cepheids in galaxies with known distances (the Milky Way, the LMC, and NGC$4258$) are used with equation \ref{eq: ceph period magnitude} to determine the parameter $M_{\rm Ceph}$, which, once determined, turns equation \ref{eq: ceph period magnitude} into a prediction for distance modulus $\mu$ for the SNIa host galaxies. Combined with the inferred values of $m_B^0$ from the SED fitter, equations \ref{eq: supernovae magnitude} and \ref{eq: ceph period magnitude} are jointly fit to observed Cepheid and SNIa in the sample to infer $M_0^{B}$.
This value, an estimate of the peak absolute $B$-band magnitude of SNIa, can be used to anchor the empirical SNIa magnitude-redshift relationship for supernovae in the Hubble flow. For each supernova in the Hubble flow,
\begin{equation}
m_b^0 = M_b^0 + 5 \log_{10} \left( D_{\rm lum}/{\rm Mpc}\right) + 25.
\end{equation}
When combined with equation \ref{eq: luminosity distance}, we obtain
\begin{equation}
m_b^0 - 5 \log_{10}\left({\frac{c(1+z)}{\rm km/s} \int_0^z \frac{dz'}{h(z')}}\right)= M_b^0 - 5 \log_{10} \left({\frac{H_0 {\rm Mpc}}{\rm km/s}}\right) + 25, 
\end{equation}
where the dependence on the observables $m_b^0$ and $z$ for each SNIa is moved to the left hand side of the equation. Since the right hand side is a constant, so must be the left hand side, and we can define 
\begin{align}
5 a_B &= m_B^0 - 5 \log_{10} \left(c(1+z)\int_0^z \frac{dz'}{h(z')}\right) \\
&\simeq m_B^0 - 5 \log_{10}\left\{cz \left\{1 + \frac{1}{2} \left(2 - h'\right) z - \frac{1}{6} \left(h'+h'' \right) z^2 \right\}\right\}, \nonumber
\end{align}
where the second equality follows from a third-order Taylor expansion in $f(z)=H_0 D_{\rm lum}(z)$ about $z=0$, $h'$ and $h''$ are successive derivatives of $h(z)$ evaluated at $z = 0$, and $c$ is in km/s. The values of $h'$ and $h''$ are estimated from measurements of $m_B^0$ and $z$ from $740$ SNIa at redshifts $z \lesssim 1$ \citep{betoule_improved_2014}, after which $a_B$ is determined from $217$ observed SNIa at redshifts $0.023 < z < 0.15$ to be $a_B = 0.71273 \pm 0.00176$ (\Riess{}). 
In turn, a local determination of $H_0$ is then given by
\begin{equation}
\label{eq : H0}
H_0 = 10^{0.2 M_b^0 + a_B + 5} \left[\frac{\rm km/s}{\rm Mpc}\right].
\end{equation}

There are four broad sources of potential systematic effects in the above analysis. In addition to potential systematics at each of the three steps of the ladder, there is an additional systematic in the analysis that leads to equation \ref{eq : H0}. This may include fast-oscillating effects in the expansion for $f(z)$ above, which is not expected in \lcdm, as well as local effects (like a local void) that impede the validity of the luminosity distance calculation in equation \ref{eq: luminosity distance}. This latter effect has been considered in e.g. \citep{marra_cosmic_2013}, who find local variation adds (or subtracts) a mean square error of around $1$ km/s/Mpc to the underlying value of $H_0$. To correct for this, \Riess{} empirically adjust the measured supernova redshifts for expected flows that trace the underlying distribution, and claim a residual $0.4\%$ uncertainty. 

\section{Expanding the Variable Star Treatment}
\label{sec: new analysis}
In this section we address potential systematics of the Cepheid treatment of equation \ref{eq: ceph period magnitude}. A more general formulation of the relationship between Cepheid measurables is
\begin{equation}
\label{eq: general Cepheid magnitude}
m_{H,i}- R_HE(V-I) = f\left(\log P_i, \log [O/H]_i, (V-I)_i^0\right) + \mu + \eta_i,
\end{equation}
where $f(\cdot)$ is a predictor of the absolute magnitude $M_i$ of a Cepheid, $R_H$ is the total-to-selective extinction in $H$ band, and $\eta_i$ is the nondeterministic contribution to Cepheid magnitude, assumed by \Riess{} to be drawn from a normal distribution of width $\sigma = 0.08$mag. The basic task is to infer $\mu$ from the Cepheid properties $m_{H_i}$, $\log P_i$, $\log [O/H]_i$, intrinsic color $(V-I)^0$, and selective extinction $E(V-I)_i$ due to dust along the line of sight. 

A primary complication in the above is that the intrinsic $(V-I)^0$ is unknown for most Cepheids in the \Riess{} sample. This issue can be sidestepped under the assumed linear relationship of equation \ref{eq: ceph period magnitude} if one assumes a uniform distribution of underlying intrinsic color between anchor and host galaxies. The observed color $V-I$ includes both intrinsic color $(V-I)^0$ and extinction along the line of sight $R_H E(V-I)$. If one makes the replacement $E(V-I)^0 \rightarrow V-I$ (as is done in \Riess{}), this leads to a systematic bias $R_H (V-I)^0$. In the case of a uniform distribution of intrinsic color along the sample, this bias is absorbed into the intercept term $M_{\rm Ceph}$ in equation \ref{eq: ceph period magnitude}. For non-linear treatments, however, or in cases where the underlying distribution of $(V-I)^0$ has dependence on metallicity, period, or other variables, this bias is not uniform over galaxies or cannot be absorbed into a simple intercept term, and potential difference in this bias between anchor galaxies with known distances and SNIa host galaxies will lead to a bias in the resulting $H_0$ determination. A first step in generalizing the linear treatment of equation \ref{eq: ceph period magnitude} to the more flexible relationship in equation \ref{eq: general Cepheid magnitude} is to gain an understanding of Cepheid intrinsic color dependencies.

\subsection{Estimation of Intrinsic Color}

\begin{figure}
\begin{centering}
\includegraphics[scale=0.48]{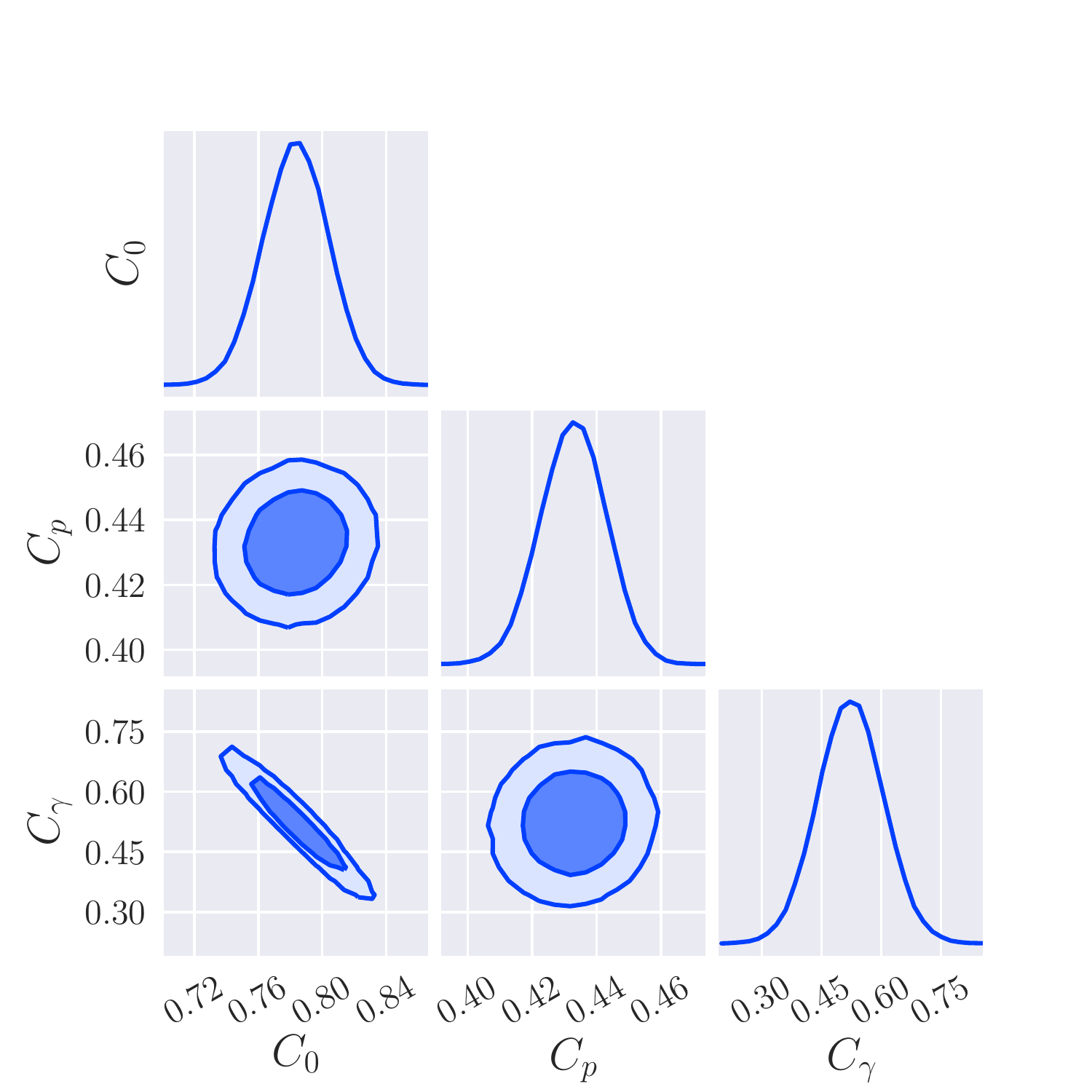}
\caption{
\label{fig: color constraints}
The constraints on the intrinsic color parameters $C_0$, $C_p$, and $C_\gamma$ of equation \ref{eq: color model} assuming the Cepheid period-magnitude relationship of equation \ref{eq: ceph period magnitude} used in \Riess{}, but with intrinsic color $(V-I)^0$ instead of the total color $(V-I)$. The posterior is jointly constrained by the Cepheid sample and our intrinsic color and period likelihood as described in the text. The degeneracy between $C_0$ and $C_\gamma$ is due to the limited metallicity information from the external color data of \citet{sandage_new_2004,tammann_new_2003}, which only contains information from LMC and Milky Way Cepheids.
}
\end{centering}
\end{figure}

Both experimental \citep{tammann_new_2003,sandage_new_2004,sandage_new_2009} and theoretical \citep{bono_theoretical_1999,ngeow_theoretical_2012} work point to possible dependence of Cepheid intrinsic color on both Cepheid period and metallicity. To model these dependencies, we introduce a linear parameterization of the intrinsic color in terms of these other Cepheid observables:
\begin{equation}
\label{eq: color model}
(V-I)^0 = C_0 + C_p (\log P -1) - C_\gamma \Delta \log[O/H].
\end{equation}
Though the metallicity dependence in particular is theoretically not expected to be linear in general \citep{ngeow_theoretical_2012}, both the relatively small range of metallicity in the sample and the dearth of experimental constraints on metallicity dependence argue for a linear interpretation. 

The \Riess{} data alone are not enough to appreciably constrain the parameters of equation \ref{eq: color model}. We therefore use additional data to constrain this equation. Estimates of intrinsic Cepheid color have been made through photometry of adjacent red clump stars \citep{udalski_optical_1999}. Relevant to the \Riess{} sample is Cepheid intrinsic color information estimated by measurements of line-of-sight dust extinction for Cepheids in the LMC \citep{sandage_new_2004} and the Milky Way \citep{tammann_new_2003}. As shown in Fig.~1b of \citet{sandage_new_2004}, these colors are well fit (up to the known width of the instability strip) by the mean relations
\begin{align}
\label{eq: intrinsic color}
 (\overline{V-I})^0 & = 0.315 \log P  + 0.695, \, \left(\text{LMC,\,} P > 10 {\rm\, days}\right) \\ \nonumber
 		 & = 0.160 \log P  + 0.661, \, \left(\text {LMC, }P < 10 {\rm \, days}\right)\\
 		 & = 0.256 \log P  + 0.497, \, \left(\text {Milky Way} \right)\nonumber,
\end{align}
which we use instead of the individual inferred intrinsic color for each sample to set constraints on $C_0$, $C_p$, and $C_\gamma$. Specifically, we fold the above predictions of equation \ref{eq: intrinsic color} into the likelihood by assuming the intrinsic color of Cepheids in the LMC and Milky Way are given by 
\begin{equation}
(\overline{V-I})^0_i = (\overline{V-I})^0 (P_i, {\rm host}_i) + \epsilon_i,
\end{equation}
with the $(\overline{V-I})^0 (P_i, {\rm host}_i)$ from equation \ref{eq: intrinsic color} and $\epsilon_i \sim \mathcal{N}(0, \sigma_\mu^2)$, and $\sigma_\mu = 0.08$ set by the width of the Cepheid instability strip. 
The final constraints on the parameters of equation \ref{eq: color model}, shown in Fig.~\ref{fig: color constraints}, are joint constraints from the above model and the Cepheid photometry of \Riess{}. In particular, $C_\gamma$, whose constraint is weak and heavily degenerate with $C_0$, is constrained mostly through metallicity trends in the observed $V-I$ of the \Riess{} sample.

Because detailed color information on Cepheids only exists for two galaxies (and therefore two metallicities), the constraints on $(V-I)^0$ for Cepheids in the sample under the model of equation \ref{eq: color model} are not expected to be precise predictions of true intrinsic color. Instead, this treatment both allows us to gain an understanding of the typical amount of reddening due to extinction $E(V-I)$, as well as allow for some variation of the distribution of intrinsic color within the sample. As explained above in section \ref{sec: R16 rehash}, this gives us the freedom to relax the linear and constant color treatment of equation \ref{eq: ceph period magnitude} done in \Riess{}, as we do below.

\subsection{Nonlinear Relationships and Internal Consistency of the Sample}
\label{sec: populations}
With the above parameterization of Cepheid intrinsic color, we can move to nonlinear parameterizations of Cepheid magnitude in equation \ref{eq: general Cepheid magnitude}. One approach is to look for second- and higher-order dependencies on the Cepheid observables. However, in the absence of strong theoretical reasons to expect a particular functional form for $f(\cdot)$, we instead adopt a less stringently parameterized approach based on Gaussian Mixture clustering of Cepheids in the period-color plane.\footnote{including metallicity does not qualitatively change the results, and leads to a significant increase in interpretive complexity.} We choose a number of clusters for $n$ Cepheids in this plane according to the Bayesian Information Criterion, 
\begin{equation}
{\rm BIC} = k \ln n - \ln {P}(d|\hat{\theta}),
\end{equation}
whose minimum over $k = 5N$ degrees of freedom over $N$ clusters corresponds to the number of clusters $N$ that maximize the likelihood of the data in the case where the true underlying distribution of Cepheids is an additive mixture of Gaussian components, and where $P(d|\hat{\theta})$ is the probability of the data under the mixture model with best fit values $\hat{\theta}$ to the means and covariances. 

\begin{figure}
\begin{centering}
\includegraphics[scale=0.4]{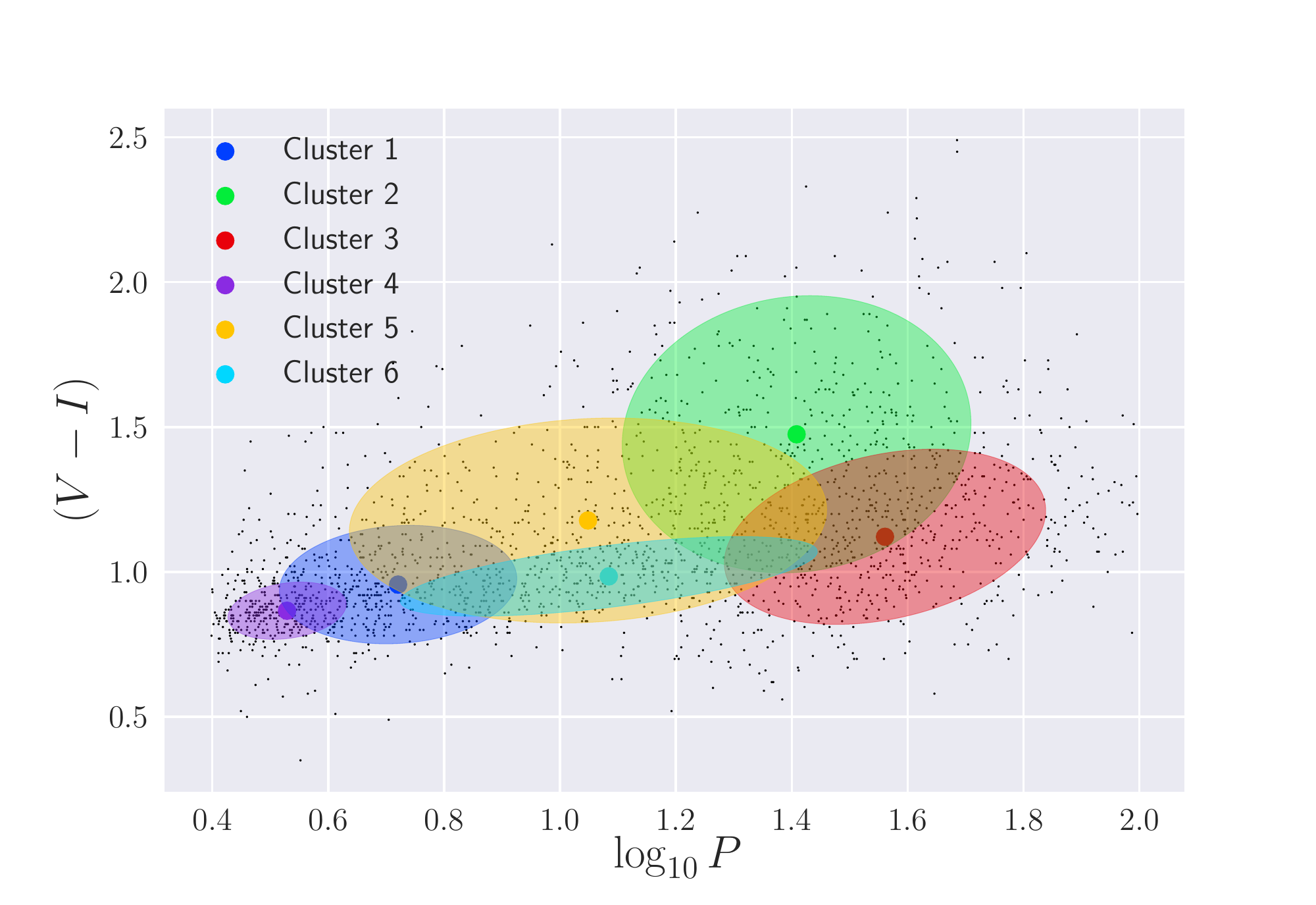}
\caption{
\label{fig: clusters}
The $68\%$ confidence regions for the six clusters that maximize the BIC criterion in period-color space. The significant overlap in cluster support leads us to adopt a weighted mixed-cluster treatment of individual Cepheids, where each Cepheid is assigned a weight in each cluster proportional to the mixture model probability of residing in that cluster. 
}
\end{centering}
\end{figure}

\begin{figure}
\begin{centering}
\includegraphics[scale=0.4]{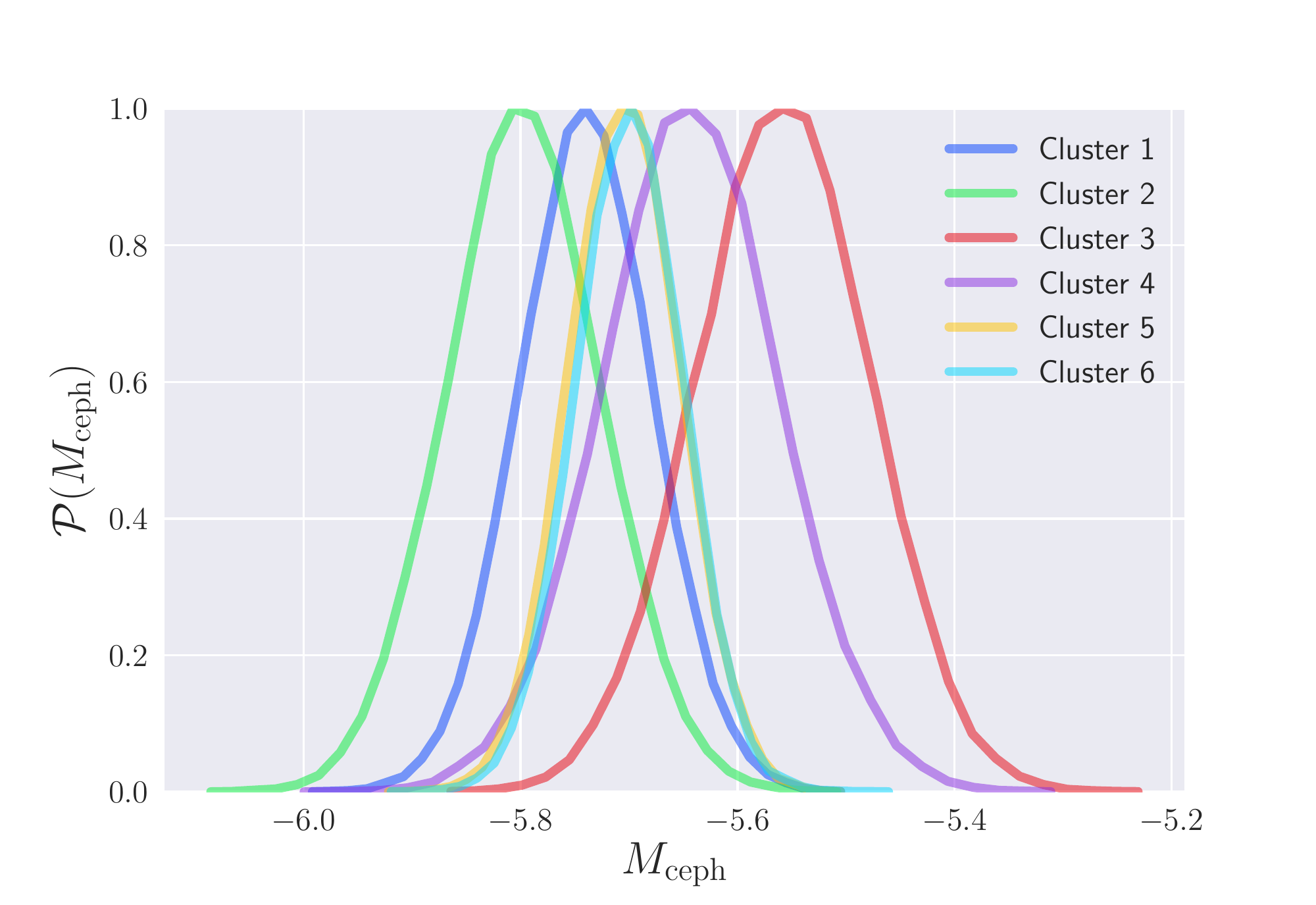}
\caption{
\label{fig: M_ceph constraints}
The posterior distribution of $M_{\rm Ceph}$ for each of the $6$ Cepheid clusters depicted in figure \ref{fig: clusters}. The constraints on $M_{\rm Ceph}$ in each cluster are consistent at the level of about $1 \sigma$ or less, as are the constraints when the results are propagated to inferences on $H_0$. The weight of the global $H_0$ constraint comes from the intermediate-period Cepheids represented by clusters $5$ and $6$ which share significant support. These clusters contain a large overlap of both host and anchor Cepheids, which reduces uncertainties from extrapolation.
}
\end{centering}
\end{figure}

This criterion selects between four and six clusters, whose distribution in period-color space (in the case of our fiducial choice of 6) is shown in Fig.~\ref{fig: clusters}. Due to the overlap in support these clusters show, rather than assigning each Cepheid a definitive cluster membership, we instead assign each Cepheid a weight in each cluster proportional to the relative probabilities of the Cepheid belonging in each cluster according to the mixture model. The sixth cluster, shown in light blue in Fig.~\ref{fig: clusters}, is of particular interest--it heavily weights intermediate-period clusters in a narrow `main sequence' range of total $V-I$. Cepheids in this region provide the most robust constraint on $H_0$ (as discussed in section \ref{sec: results}), which is due to the large overlap between Cepheids in SNIa host galaxies and Cepheids in anchor galaxies with known geometric distances in this range. On the other hand, Cluster 4 (purple in Fig.~\ref{fig: clusters}) only carries significant support from a few Cepheids outside the LMC, and clusters 2 and 3 (green and red respectively) draws support from relatively few anchor Cepheids. As a result, these clusters contribute much less to the global $H_0$ constraint. The inference on $H_0$ from the different populations, driven by the Cepheid magnitude estimates in Fig.~\ref{fig: M_ceph constraints}, are all self-consistent, with the largest discrepancy (between $H_0$ inferred from clusters 3 and 6) being seperated by $ \lesssim 1\sigma$.

Each cluster in the mixture model is then fit with an individual, linear period-luminosity relationship as expressed in equation \ref{eq: ceph period magnitude}, leading to six different estimates of the parameters $M_{\rm Ceph}$, $b$, and $\gamma$ whose support comes from Cepheids in different parts of the period-color phase space. This soft-boundary independent-constraint approach allows potential nonlinear information to be captured without the introduction of hard cuts in one or more dimensions of Cepheid observables--as in, for example, the hard $P = 10$ days break used in R$16$ and elsewhere.

The distributions of the $M_{\rm Ceph}$ constraints from each of the $6$ clusters is shown in figure \ref{fig: M_ceph constraints}, where larger values of $M_{\rm Ceph}$ equate to larger inferred peak SNIa brightness $M_{b}^0$ and therefore larger $H_0$. With the typical values for the Cepheid period-magnitude relationship, the value of $M_{\rm Ceph} = -5.8$ preferred by highly reddened, high period Cepheids corresponds to $H_0 \simeq 68~{\rm km/s/Mpc}$, in line with Planck measurements, though the $M_{\rm Ceph} = -5.5$ preferred by slightly reddened high period Cepheids corresponds to $H_0 \simeq 79~{\rm km/s/Mpc}$, a value in unambiguous disagreement with the CMB inference. The global constraint with $\hat{H}_0 \simeq 72~{\rm km/s/Mpc}$ is driven by the inference from `main sequence' Cepheids with moderate period and typical reddening most captured by cluster $6$.

\subsection{Variability of Total-to-Selective Extinction over Lines of Sight}
The treatment so far assumes a fixed value of total-to-selective extinction, where
\begin{equation}
R_H \equiv \frac{A(H)}{E(V-I)}
\end{equation}
is held constant through the entire analysis, and in fact is set equal to the externally-determined value in the Milky Way, $R_H = 0.39$. Reddening from dust outside the Milky Way is not well studied \citep{fitzpatrick_correcting_1999}, and uncertainties in the reddening in host galaxies is a major source of uncertainty in analysis at shorter wavelengths \citep{Freedman}. While the use of H-band photometry by \Riess{} lowers the sensitivity enormously, there is still the possibility that a color bias that preferentially reddens anchor galaxies over Cepheid host galaxies would mimic the effects of a larger luminosity distance to hosts relative to anchors, and therefore larger inferred $H_0$. 

\begin{figure}
\begin{centering}
\includegraphics[scale=0.4]{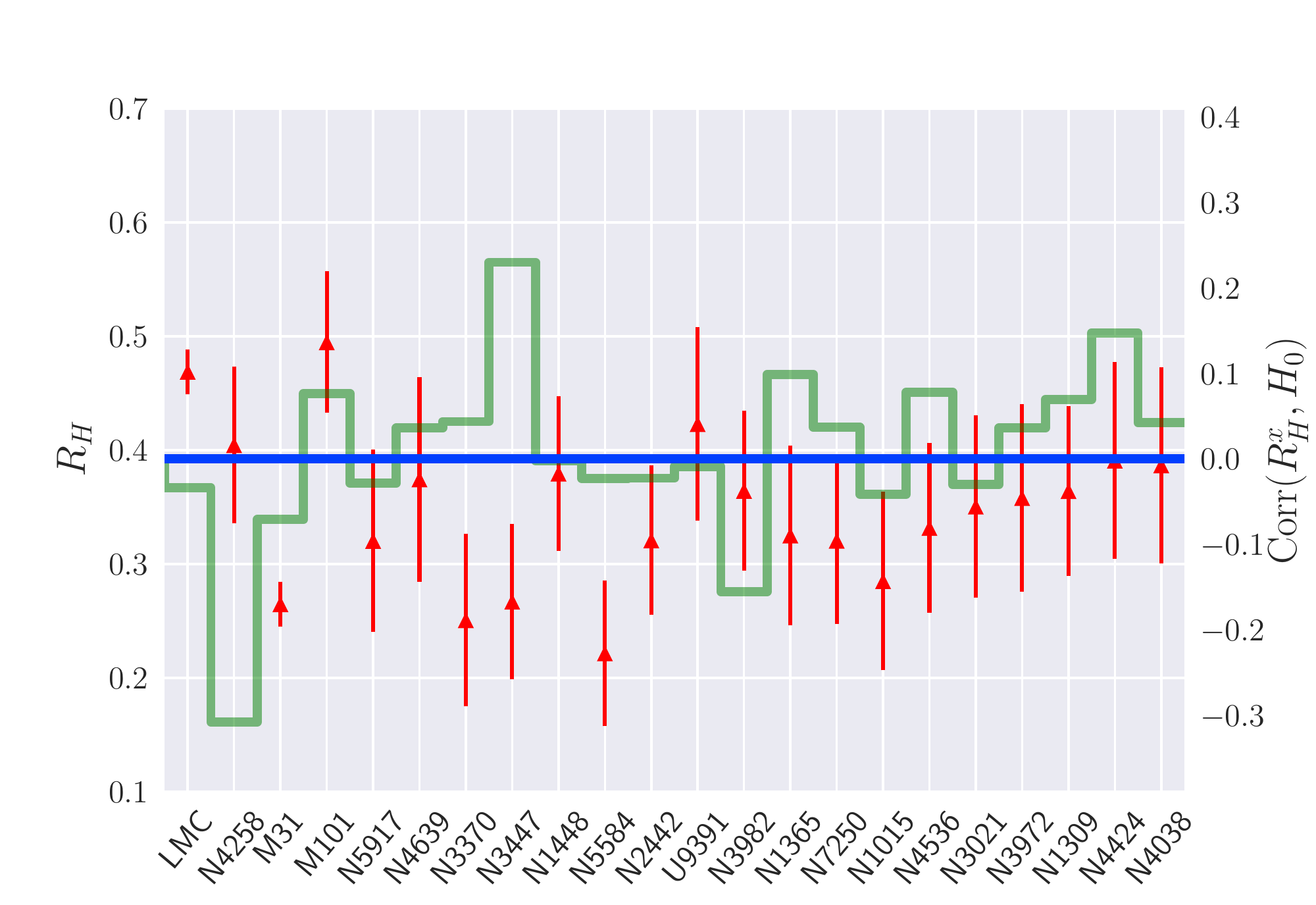}
\caption{
\label{fig: color variation}
The contraint on the value of $R_H$ for each galaxy in the sample, under a wide prior of $R_H = 0.39 \pm 0.1$ to regularize hosts with insufficient information for a strong determination. Overlaid is the value of the correlation of $H_0$ with the value of $R_H$ in each galaxy; a measurement of $R_H$ in fields with significant correlation will have the most effect on $H_0$.
}
\end{centering}
\end{figure}

The existence of multiple Cepheids in each host in principle allows a determination of a mean excursion $\delta R_H^\alpha$ from the Milky Way value of $R_H = 0.39$ \cite{fitzpatrick_correcting_1999} in each galaxy $\alpha$. Fig.~\ref{fig: color variation} shows the constraints on $R_H$ for each galaxy in the sample. In practice, the Cepheid counts in most galaxies (apart from the LMC and M31) are insufficient to make a completely data-driven inference on $R_H^\alpha$; constraints and correlation shown are from a wide prior of $R_H^\alpha = 0.39 \pm 0.1$ \citep{fitzpatrick_correcting_1999}. As a measure of the importance of the determination of $R_H$ in each host, the correlation of $\delta R_H$ in each field with $H_0$ under the model of equation \ref{eq: general Cepheid magnitude} is overlaid. 

Since M31 neither has a geometric measure nor hosts a supernova (and is therefore used solely to constrain the nuisance parameters of equation \ref{eq: general Cepheid magnitude}), it is unsurprising that the preference for $R_H < 0.39$ has no noticeable effect on $H_0$. Similarly, the evidence of $R_H > 0.39$ along the line of sight to the LMC Cepheids does little because the LMC is both host to much of the sample and preferentially hosts low-period Cepheids, which allows the slope of the period-luminosity relationship to adjust to accomodate the increased value of $R_H^{\rm LMC}$. 

Of particular importance is the value of $R_H$ in the direction of NGC$4258$, which plays an outsized role due to the fact the Cepheid sample there most accurately matches the period-color distribution of Cepheids in supernova host galaxies. With the limited data available, $R_H^{{\rm NGC}4258}$ is compatible with the assumed Milky Way value, though more information on dust extinction in this galaxy may shed additional light. It is worth noting, however, that to completely explain the tension between $Planck$-derived and Cepheid-derived Hubble measurements, we would need $ R_H \simeq 0.6$ along the line-of-sight to NGC$4258$: an extremely unlikely excursion, and one in no way favored by the local distance data.

\section{Results}
\label{sec: results} 

\begin{figure}
\begin{centering}
\includegraphics[scale=0.4]{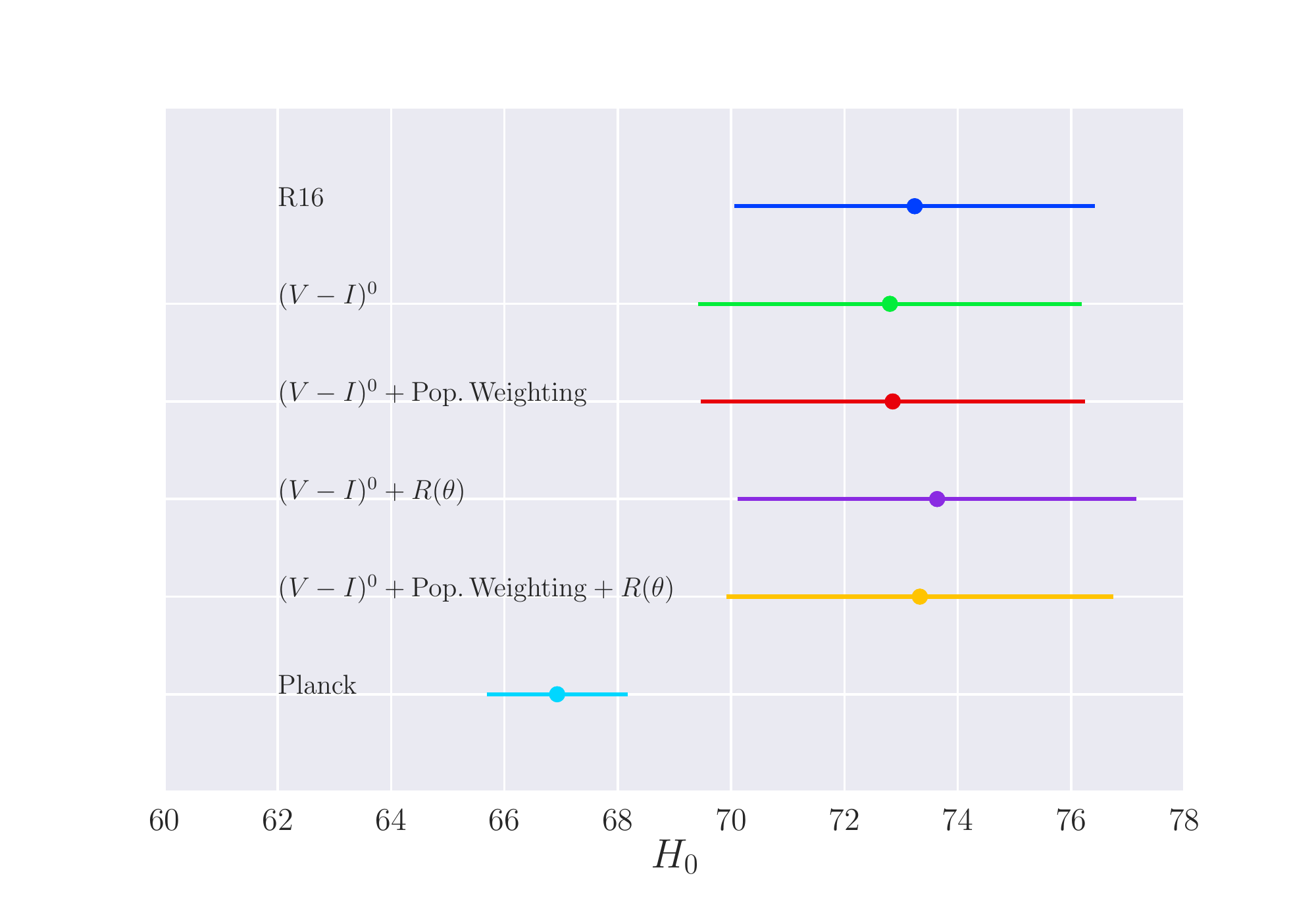}
\caption{
\label{fig: H_0 constraints}
The contraints on $H_0$ under the various extended models of section \ref{sec: new analysis} compared to the baseline model of \Riess{}. The value of $H_0$ shows remarkable consistency across all models, and the constraint (both in expectation and variance) shows tremendous robustness to particular assumptions in the Cepheid magnitude model of equation \ref{eq: general Cepheid magnitude}.
}
\end{centering}
\end{figure}

\begin{figure}
\begin{centering}
\includegraphics[scale=0.4]{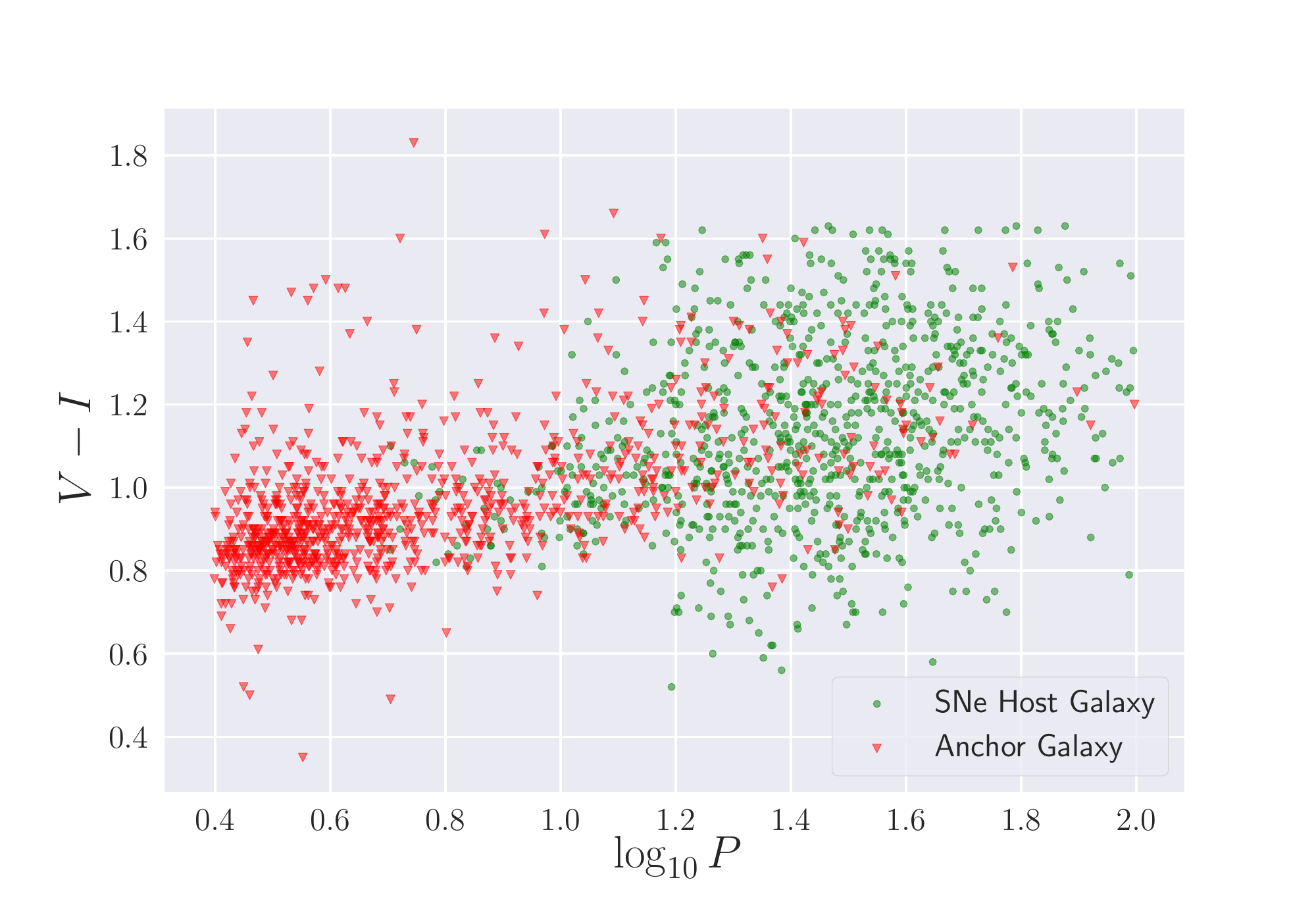}
\caption{
\label{fig: train vs test}
The distribution in period-reddening space of Cepheids in SNe host galaxies (green circles) vs anchor galaxies (red triangles). While in general observations have detected higher period Cepheids in host galaxies and lower period Cepheids in anchor galaxies, in the range of periods from $0.8 < \log_{10} P < 1.4$, which dominates the constraint, there is significant overlap.
}
\end{centering}
\end{figure}

The constraints for the models described in section \ref{sec: new analysis} are shown in Fig.~\ref{fig: H_0 constraints}, with the most general model ($(V-I)^0 + {\rm Pop. \, Weighting} +R(\theta)$, which includes all generalizations detailed in section \ref{sec: new analysis}) giving a value of $H_0 = 73.3 \pm 1.7~{\rm km/s/Mpc}$. The value (and precision of determination) of $H_0$ shows remarkable consistency across all models considered. The root cause of this consistency is that very little extrapolation is actually necessary to map a host Cepheid to an anchor Cepheid with known absolute magnitude for the `main sequence' Cepheids that drive the constraint (Cluster 6 in Fig.\ref{fig: clusters}). The results of the previous section can all be viewed as a special case of this fact--no generalization of the model relating anchor and host Cepheid magnitude measurements can introduce significant bias in the $H_0$ inference. The approaches of sections \ref{sec: R16 rehash} and \ref{sec: new analysis} are all examples of an attempt to infer intrinsic magnitudes of Cepheids in SNIa host galaxies (shown in green in Fig.~\ref{fig: train vs test} through an interpolation of Cepheids in anchor galaxies (shown in red in Fig ~\ref{fig: train vs test}) with known geometric distances, and therefore known intrinsic magnitudes, through an interpolative model. It is instructive to compare the results of \Riess{} to an `interpolation-free' model, where the absolute magnitude of a Cepheid is simply given as the absolute magnitude of its nearest neighbor for some suitably defined metric on the space of observed Cepheid features.

\begin{figure}
\begin{centering}
\includegraphics[scale=0.4]{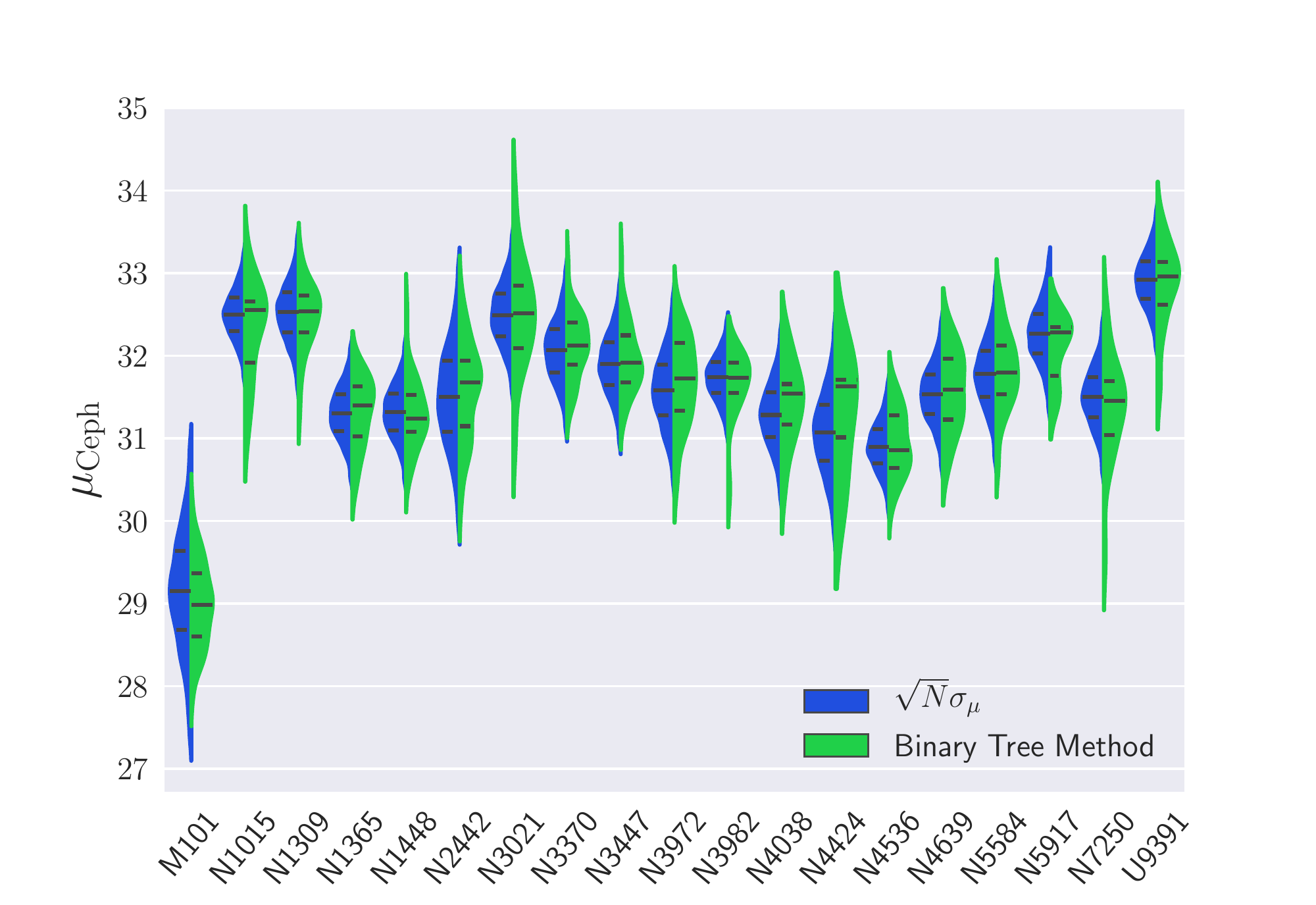}
\caption{
\label{fig: CART constraint}
A plot of the distribution of point estimates of the distance modulus from each Cepheid to its host galaxy from a `model free' approach to assigning absolute magnitudes to host Cepheids through assigning them the absolute magnitude of their nearest neighbor in the anchor sample. Lines show the quartiles of the distributions. Here distance is determined through the recursive tree algorithm described in the text, where the Cepheid period/color/metallicity feature space is iteratively divided along an axis to minimize the resulting sum-of-variance in the subpopulations, until each population consists of a single Cepheid in an anchor galaxy and its neighbor Cepheids in a host galaxy. The distribution of these maximum likelihood estimates is compared to the expected distribution of maximum-likelihood estimates under the model of \Riess{} and the assumption of independent and equal errors on the estimates, given by $\sqrt{N}\sigma_{\mu}$, where $N$ is the number of Cepheids in the host and $\sigma_{\mu}$ is the error on the Cepheid distance determination from Table 5 of \Riess{}. Approximately $5\%$ of the distance modulus determinations from this procedure result in clear outliers (because the Cepheid in question is far removed from a Cepheid with known absolute magnitude); these are suppressed in the plot.
}
\end{centering}
\end{figure}

Such a result is shown in Fig.~\ref{fig: CART constraint}. To define neighboring Cepheids, the space of Cepheid $\left\{ \log P, V-I, {\rm and } \log [O/H] \right\}$ is segmented through iteratively partitioning the population of anchor Cepheids along an axis at a location that minimizes the sum of squared residuals to the means of the resulting subpopulations, creating a binary decision tree branching over cuts along the 3 axes of Cepheid observables. This procedure continues until all leaves of the tree contain a single anchor Cepheid. Each Cepheid in the host Galaxy sample is then propogated forwards through these branches, until it terminates in a leaf of the tree. These host Cepheids are then assigned the absolute magnitude of the `neighboring' anchor Cepheid it shares a leaf with; assigning an absolute magnitude for each Cepheid in each host. Combined with their best-estimated observed magnitude, these absolute magnitudes give a point estimate of the distance modulus to the host galaxy and its hosted SNIa. 

For comparison, this distribution is shown against the expected variation in point estimates under the model of \Riess{}. Under the approximatley valid assumption of independent and equal errors, this variation corresponds to a Gaussian centered at the predicted value of $\mu^\alpha$ for galaxy $\alpha$ and with width given by $\sqrt{N}\sigma_\mu$ (where $N$ is the number of Cepheids in the host galaxy). A shift in $H_0$ sufficient to relieve the tension ($\Delta H_0/H_0 \simeq 10\%$) requires a change in distance modulus of $\Delta \mu \simeq 0.5$; this is well above the shift in the means of the distribution of point estimates when switching between the models observed in Fig.~\ref{fig: CART constraint}.

\section{Conclusion}
\label{sec: conclusion}
This paper investigates the relaxation of assumptions in modeling Cepheid apparent magnitudes as a way to reduce the tension between the CMB inference and the distance ladder determination. Several extensions were considered to take into account potential dependence on differential bias in sampling between anchor Cepheids in galaxies with known geometric distance and host Cepheids in galaxies with SNIa. None of the extensions appreciably reduce the tension. Indeed, the inference is essentially independent of modeling choices, due to the proximity in the sample of Cepheids from each population in the feature space of Cepheid period, color, and metallicity. While this still leaves the possibility of internal tensions between inferences drawn from different regions of the Cepheid feature space (meaning selection choices can affect the inference), our six-cluster analysis in section \ref{sec: populations} indicates this is not the case. There we found that constraints on $H_0$ conditioned on Cepheids in different regions of the Cepheid feature space are internally consistent, and the global constraint is broadly driven by intermediate period Cepheids with typical values of $V-I$.

Differences in environmental factors between galaxies can in principle also explain the tension. The \Riess{} analysis relies on a differential measurement between Cepheid samples in host and anchor galaxies. A bias in the $H_0$ inference can occur if these samples are not drawn from the same population--either because the Cepheid distribution itself varies between host and anchor galaxies or because sampling or measurement introduces differential bias between the two. Of particular interest is the anchor galaxy NGC$4258$, which plays an outsized role in more generalized models due to being the dominant source of the intermediate-to-high-period Cepheids most similar to the Cepheids measured in supernova host galaxies. Generalizing the Cepheid period-magnitude relationship can in principle decouple low-period Cepheids from the bulk of the SNIA host galaxy Cepheids with higher magnitudes, increasing the importance of the NGC$4258$ distance measure in constraining $H_0$, and motivating a search for potential bias in Cepheid modeling in that galaxy. Dependence of $R_H$ along the line of sight to NGC$4258$, as shown in the correlations in Fig.~\ref{fig: color variation}, is one source of potential differential bias; however, the magnitude of the effect, even for large variations in $R_H$, is too small to explain the tension. 

Other steps in the distance ladder may also contribute systematic effects, which are not considered here. From selective changes in the analysis pipeline that leads to the fiducial result, \Riess{} estimate the uncertainty from these effects (including in the Cepheid period-magnitude relationship) at around $\sigma = 1~{\rm km/s/Mpc}$. How this estimate generalizes to the treatments in section \ref{sec: new analysis} is unclear; we opt to quote uncertainties without these systematic effects, and note only their presence. These include potential biases in the parallax measurements of Milky Way Cepheids \citep{Benedict:2006cp} or the geometric measures to the LMC \citep{fitzpatrick_fundamental_2000} or NGC$4258$ \citep{Humphreys:2013eja}, photometric bias in the Cepheid magnitude measurements, and biases either experimental or real in the low-redshift SNIa with companion Cepheids used to calibrate the SNIa magnitude-redshift relation. It is also possible that errors affecting the shape of the magnitude-redshift relation can also lead to changes in $H_0$ through biasing the constraint on the intercept $a_B$; however, the consistency of the SNIa magnitude-redshift relation to the CMB inference when anchored to CMB or BAO derived distance scales \citep{Alam:2016hwk} argues against such a bias being the culprit for the apparent tension. 

All told, our analyses show remarkable consistency in the Cepheid-calibrated $H_0$ determination to modeling choices in the distribution of Cepheids, and effectively rule out bias in Cepheid photometric modeling as a means of alleviating the tension between distance ladder and CMB-derived $\Lambda$CDM inferences of $H_0$. Absent unaccounted confirmation bias, the presence of three self-consistent geometric anchors in the LMC eclipsing binaries, the NGC$4258$ water maser, and Milky Way Cepheid parallaxes (plus a distance to M$31$ consistent with the others \citep{riess_private_2016} but unused in the analysis) argue further against systematic bias in determining the geometric distances that anchor the Cepheid magnitude relationship. A recent analysis of $212$ Cepheids in the Milky Way from $Gaia$ by \citet{casertano_test_2017} independently anchors the Cepheids and finds $0.3\%$ agreement with the \Riess{} determination of $H_0$, further buttressing the first rung of the distance ladder.

It remains a possibility that there is an unaccounted-for systematic in Cepheid photometry or in the analysis of the SNIa in the nearby host galaxies, neither of which have we revisited.
Independent analyses of these parts of the inference chain would be very valuable. Complementary local probes of cosmic expansion with independent sources of systematic errors, such as $H_0$ inferences from lensed quasar time delays \citep{suyu_h0licow_2017}, hold perhaps even greater promise to settling concerns over systematic bias; progress in increasing the precision of alternative measures may ultimately prove the arbiter between this tension as a harbinger of new physics, or simply a statistical or systematic artifact.

\section{Acknowledgements: }
We thank Adam Riess for assistance with reproduction of the R16 analysis and comments on a draft manuscript and James Aguirre, Tucker Jones, Matt Richter, Abhijit Saha, Kendrick Smith, and Stefano Valenti for useful conversations.
\label{lastpage}

\bibliographystyle{mnras}
\bibliography{paper}

\end{document}